\input phyzzx
\input phyzzx.plus
\overfullrule=0pt

\def\Tr{\mathop{\rm Tr}\nolimits}

\def\Pf{\mathop{\rm Pf}\nolimits}
\REF\LUS{M.~L\"uscher, Nucl.\ Phys.\ B 549 (1999) 295.}
\REF\SUZ{H.~Suzuki, Prog.\ Theor.\ Phys.\ 101 (1999) 1147.}
\REF\LUSC{M.~L\"uscher, Nucl.\ Phys.\ B 568 (2000) 162;
Nucl.\ Phys.\ Proc.\ Suppl.\ 83 (2000) 34.}
\REF\SUZU{H.~Suzuki, Nucl.\ Phys.\ B 585 (2000) 471.}
\REF\KIK{Y.~Kikukawa, Y.~Nakayama, hep-lat/0005015.}
\REF\LUSCH{M.~L\"uscher, JHEP 06 (2000) 028.}
\REF\LUSCHE{M.~L\"uscher, Nucl.\ Phys.\  B 538 (1999) 515;
T.~Fujiwara, H.~Suzuki, K.~Wu, Nucl.\ Phys.\ B 569 (2000) 643;
Phys.\ Lett.\ B 463 (1999) 63; hep-lat/9910030.}
\REF\NEU{H.~Neuberger, Phys.\ Rev.\ D 59 (1999) 085006;
D.H.~Adams, hep-lat/9910036, to appear on Phys.\ Rev.\ Lett.;
Nucl.\ Phys.\ B 589 (2000) 633.}
\REF\BAR{O.~B\"ar, I.~Campos, Nucl.\ Phys.\ Proc.\ Suppl.\ 83 (2000)
594; Nucl.\ Phys.\ B 581 (2000) 499.}
\REF\WIT{E.~Witten, Phys.\ Lett.\ 117 B (1982) 324;
S.~Elitzur, V.P.~Nair, Nucl.\ Phys.\ B 243 (1984) 205.}
\REF\NEUB{H.~Neuberger, Phys.\ Lett.\ B 437 (1998) 117.}
\REF\ALV{L.~Alvarez-Gaum\'e, E.~Witten, Nucl.\ Phys.\ B 234 (1984)
269.}
\REF\HSU{S.D.H.~Hsu, Mod.\ Phys.\ Lett.\ A 13 (1998) 673.}
\REF\KAP{D.B.~Kaplan, M.~Schmaltz, Chin.\ J. Phys.\ 38 (2000) 543;
G.T.~Fleming, J.B.~Kogut, P.M.~Vranas, hep-lat/0008009.}
\REF\KAPL{D.B.~Kaplan, Phys.\ Lett.\ B 288 (1992) 342;
Nucl.\ Phys.\ Proc.\ Suppl.\ 30 (1993) 597;
Y.~Shamir, Nucl.\ Phys.\ B 406 (1993) 90;
Nucl.\ Phys.\ Proc.\ Suppl.\ 47 (1996) 212;
V.~Furman, Y.~Shamir, Nucl.\ Phys.\ B 439 (1995) 54.}
\REF\CUR{G.~Curci, G.~Veneziano, Nucl.\ Phys.\ B 292 (1987) 555;
I.~Montvay, Nucl.\ Phys.\ B 466 (1996) 259;
Nucl.\ Phys.\ Proc.\ Suppl.\ 53 (1997) 853; 63 (1998) 108; 83 (2000)
188;
G.~Koutsoumbas, I.~Montvay, Phys.\ Lett.\ B 398 (1997) 130;
G.~Koutsoumbas, I.~Montvay, A.~Pap, K.~Spanderen, T.~Talkenberger,
J.~Westphalen, Nucl.\ Phys.\ Proc.\ Suppl.\ 63 (1998) 727;
DESY-Munster Collaboration (R.~Kirchner et al.),
Nucl.\ Phys.\ Proc.\ Suppl.\ 73 (1999) 828; Phys.\ Lett.\ B 446 (1999)
209;
DESY-Munster Collaboration (I.~Campos et al.), Eur.\ Phys.\  J.\ C 11
(1999) 507;
DESY-Munster Collaboration (A.~Fao et al.),
Nucl.\ Phys.\ Proc.\ Suppl.\ 83 (2000) 661; 83 (2000) 670;
A.~Donini, M.~Guagnelli, Phys.\ Lett.\ B 383 (1996) 301;
A.~Donini, M.~Guagnelli, P.~Hern\'andez, A.~Vladikas,
Nucl.\ Phys.\ Proc.\ Suppl.\ 63 (1998) 718; Nucl.\ Phys.\ B 523 (1998)
529;
A.~Donini, E.~Gabrielli, M.B.~Gavela, Nucl.\ Phys.\ Proc.\ Suppl.\ 73
(1999) 721; Nucl.\ Phys.\ B 546 (1999) 119.}
\REF\HUE{P.~Huet, R.~Narayanan, H.~Neuberger, Phys.\ Lett.\ B 380
(1996) 291;
S.~Aoki, K.~Nagai, S.V.~Zenkin, Nucl.\ Phys.\ B 508 (1997) 715;
Nucl.\ Phys.\ Proc.\ Suppl.\ 63 (1998) 602;
J.~Nishimura,
Phys.\ Lett.\ B 406 (1997) 215; Nucl.\ Phys.\ Proc.\ Suppl.\ 63 (1998)
721;
N.~Maru, J.~Nishimura, Int.\ J. Mod.\ Phys.\ A 13 (1998) 2841;
T.~Hotta, T.~Izubuchi, J.~Nishimura, Nucl.\ Phys.\ Proc.\ Suppl.\ 63
(1998) 685; Mod.\ Phys.\ Lett.\ A 13 (1998) 1667;
T.~Aoyama, Y.~Kikukawa, Phys.\ Rev.\ D 59 (1999) 054507;
U.M.~Heller, R.~Edwards, R.~Narayanan, Phys.\ Lett.\ B 438 (1998) 96;
Nucl.\ Phys.\ Proc.\ Suppl.\ 73 (1999) 497; Chin.\ J. Phys.\ 38 (2000) 594;
W.~Bietenholz, Mod.\ Phys.\ Lett.\ A 14 (1999) 51;
H.~So, N.~Ukita, Phys.\ Lett.\ B 457 (1999) 314.}
\REF\NEUBE{H.~Neuberger, Phys.\ Rev.\ D 57 (1998) 5417.}
\REF\GIN{P.H.~Ginsparg, K.G.~Wilson, Phys.\ Rev.\ D 25 (1982) 2649.}
\REF\HAS{P.~Hasenfratz, Nucl.\ Phys.\ Proc.\ Suppl.\ 63 (1998) 53;
Nucl.\ Phys.\ B 525 (1998) 401.}
\REF\NEUBER{H.~Neuberger, Phys.\ Lett.\ B 417 (1998) 141; B 427 (1998)
353.}
\REF\HASE{P.~Hasenfratz, V.~Laliena, F.~Niedermayer, Phys.\ Lett.\ B
427 (1998) 125;
M.~L\"uscher, Phys.\ Lett.\ B 428 (1998) 342.}
\REF\HER{P.~Hern\'andez, K.~Jansen, M.~L\"uscher, Nucl.\ Phys.\ B 552
(1999) 363.}
\REF\LUSCHER{M.~L\"uscher, Commun.\ Math.\ Phys.\ 85 (1982) 39;
A.V.~Phillips, D.A.~Stone, Commun.\ Math.\ Phys.\ 103 (1986) 599;
131 (1990) 255.}
\REF\FUJ{T.~Fujiwara, H.~Suzuki, K.~Wu, hep-lat/0001029.}
\REF\NIE{F.~Niedermayer, Nucl.\ Phys.\ Proc.\ Suppl.\ 73 (1999) 105;
R.~Narayanan, Phys.\ Rev.\ D 58 (1998) 097501;
Y.~Kikukawa, A.~Yamada, Nucl.\ Phys.\ B 547 (1999) 413.}
\REF\FUJI{K.~Fujikawa, Phys.\ Rev.\ D 29 (1984) 285.}
\REF\CHI{T.W.~Chiu, Phys.\ Rev.\ D 58 (1998) 074511;
K.~Fujikawa, Phys.\ Rev.\ D 60 (1999) 074505.}
\REF\NAR{R.~Narayanan and H.~Neuberger, Phys.\ Lett.\ B 302 (1993) 62;
Phys.\ Rev.\ Lett.\ 71 (1993) 3251; Nucl.\ Phys.\ B 412 (1994) 574;
B 443 (1995) 305;
R.~Narayanan, Nucl.\ Phys.\ Proc.\ Suppl.\ 34 (1994) 95;
H.~Neuberger, Nucl.\ Phys.\ Proc.\ Suppl.\ 83 (2000) 67;
S.~Randjbar-Daemi and J.~Strathdee, Phys.\ Lett.\ B 348 (1995) 543;
Nucl.\ Phys.\ B 443 (1995) 386; B 466 (1996) 335; Phys.\ Lett.\ B 402
(1997) 134.}
\Pubnum={CERN-TH/2000-290\cr IU-MSTP/43\cr}
\titlepage
\title{Real Representation in Chiral Gauge Theories on the Lattice}
\author{Hiroshi Suzuki\foot{hsuzuki@mito.ipc.ibaraki.ac.jp}}
\address{Department of Mathematical Sciences, Ibaraki University,
Mito 310-8512, Japan}
\abstract{The Weyl fermion belonging to the real representation of the
gauge group provides a simple illustrative example for L\"uscher's
gauge-invariant lattice formulation of chiral gauge theories. We can
explicitly construct the fermion integration measure globally over the
gauge-field configuration space in the arbitrary topological sector;
there is no global obstruction corresponding to the Witten anomaly. It
is shown that this Weyl formulation is equivalent to a lattice
formulation based on the Majorana (left--right-symmetric) fermion, in
which the fermion partition function is given by the Pfaffian with a
definite sign, up to physically irrelevant contact terms. This
observation suggests a natural relative normalization of the fermion
measure in different topological sectors for the Weyl fermion
belonging to the complex representation.}
\bigskip\bigskip
\noindent
PACS numbers: 11.15.Ha, 11.30.Rd\hfill\break
\endpage
\chapter{Introduction}
A general strategy to implement anomaly-free chiral gauge theories on
the lattice while preserving the {\it exact\/} gauge invariance has
emerged recently~[\LUS--\LUSCH]. In this paper, we apply this
formulation to a single Weyl fermion, which belongs to a {\it real\/}
representation\foot{This is sometimes called the real-positive
representation in the literature.} of the gauge group. Our motivation
is two fold:

In the formulation of~[\LUS,\LUSC], there are two kinds of obstruction
that prevent the gauge-invariant formulation. The first is the
{\it local\/} gauge anomaly that corresponds to the gauge anomaly in
the continuum theory, but requires a control with finite lattice
spacings~[\LUS--\LUSCHE] (see also~[\NEU]). The second is the
{\it global\/} topological obstruction~[\LUS,\LUSC,\BAR], which is a
lattice counterpart of the Witten anomaly~[\WIT].\foot{The Witten
anomaly in lattice gauge theory has been studied also from the
viewpoint of the spectral flow~[\NEUB].} The local anomaly is absent
from real representations, so we expect that global issues in the
formulation are highlighted. In fact, we can show that there is no
global obstruction for real representations and that it is always
possible to construct the gauge-invariant fermion integration measure
globally, over the gauge-field configuration space.\foot{For
{\it pseudo}-real representations of~$SU(2)$, from which the local
anomaly is also absent, it has been shown~[\BAR] that a globally
consistent definition of the fermion integration measure is
impossible.} This is the expected result from the knowledge in the
continuum theory~[\WIT,\ALV,\HSU]. We will explicitly construct such a
measure and, with that measure, we can work out all the quantities in
the formulation, including fermion expectation values in general
topological sectors. In this way, real representations provide an
illustrative example for the formulation.

Secondly, there has been a renewed interest~[\KAP] in the context of
the domain wall fermion~[\KAPL] on a lattice formulation of SUSY
Yang--Mills theories~[\CUR--\NEUBE], in which the fermion (the
gaugino) belongs to the real representation, i.e.\ the adjoint
representation. Usually such a fermion is regarded as the Majorana
fermion because either Weyl or Majorana is a matter of convention in
four-dimensional continuum theory and the latter is more symmetric
with respect to the chirality. However it is not obvious whether or
not the lattice formulation based on the Weyl fermion~[\LUS--\LUSCH]
and that based on the Majorana fermion are equivalent. We will show
that fermion expectation values in general topological sectors differ,
in the two formulations, only by contact terms that are irrelevant in
physical amplitudes. Thus they are actually physically equivalent.
This result supports the view that the framework of~[\LUS,\LUSC]
provides a unified treatment of chiral gauge theories in general. The
matching between the Weyl and the Majorana formulations moreover
suggests a natural relative normalization of the fermion integration
measure in different topological sectors for the Weyl fermion in the
{\it complex\/} representation.

We begin with recapitulating some basics of the formulation. For
unexplained notations and for more details, see~[\LUS,\LUSC]. We
assume that the lattice volume is finite throughout this paper.

\chapter{Real representations in L\"uscher's formulation}
In the formulation of~[\LUS,\LUSC], the expectation value of an
operator~${\cal O}$ in the fermion sector is defined by the path
integral
$$
   \VEV{{\cal O}}_{\rm F}=\int {\rm D}[\psi]{\rm D}[\overline\psi]\,
   {\cal O}e^{-S_{\rm F}}.
\eqn\twoxone
$$
In this paper, the fermion action is taken as\foot{The matrix $B$~is
defined by~$B=C\gamma_5$ from the charge conjugation matrix~$C$. We
take the representation of the Dirac algebra such that
$C\gamma_\mu C^{-1}=-\gamma_\mu^T=-\gamma_\mu^*$,
$C\gamma_5C^{-1}=\gamma_5^T=\gamma_5^*$, $C^\dagger C=1$ and~$C^T=-C$.
These imply $B\gamma_\mu B^{-1}=\gamma_\mu^T=\gamma_\mu^*$,
$B^\dagger B=1$ and~$B^T=-B$.}
$$
   S_{\rm F}=a^4\sum_x\biggl[
   \overline\psi(x)D\psi(x)+{1\over2}im\psi^T(x)B\psi(x)
   -{1\over2}im\overline\psi(x)B^{-1}\overline\psi^T(x)\biggr],
\eqn\twoxtwo
$$
where the Dirac operator~$D$ satisfies the Ginsparg--Wilson~(GW)
relation~[\GIN] $\gamma_5D+D\gamma_5=aD\gamma_5D$. We require that
$D$~be gauge-covariant and that it depends, locally and smoothly, on
the gauge field. Such a Dirac operator in fact exists~[\HAS,\NEUBER].
The locality and the smoothness are, however, guaranteed only in a
restricted gauge-field configuration space, as expected from the index
relation~[\HASE]. For the overlap-Dirac operator~[\NEUBER], the
sufficient condition is~[\HER]
$$
   \bigl\|1-R[U(p)]\bigr\|
   <\epsilon\qquad\hbox{for all plaquettes $p$},
\eqn\twoxthree
$$
where $R$~is the representation of the gauge group and $\epsilon$~is
any fixed positive number smaller than~$1/30$. Under this
admissibility, the gauge-field configuration space is divided into
topological sectors~[\LUSCHER,\LUS] (see also~[\FUJ]). As further
requirements, we assume the
$\gamma_5$-hermiticity~$D^\dagger=\gamma_5D\gamma_5$ and the charge
conjugation property~$D^*=BD_{R\to R^*}B^{-1}$,\foot{Throughout this
paper, the complex conjugation and the transpose operation on an
operator are defined with respect to the corresponding kernel in
position space.} where $R^*$~is the complex conjugate representation
of~$R$. For real representations, we may take
$R(T^a)^*=R(T^a)=-R(T^a)^T$, where $R(T^a)$ is the representation
matrix for the Lie algebra of the gauge group. This
implies~$D^*=BDB^{-1}$ for real representations.

In~\twoxtwo, we have introduced the ``Majorana'' mass terms to treat
topologically non-trivial sectors, in which there are zero modes of
the Dirac operator, just as easily as the vacuum sector. If one is
interested in the massless theory, it is sufficient to take the
$m\to0$~limit at the very end of calculations. The mass terms are
consistent with the Fermi statistics and, for real representations,
gauge-invariant.

The chirality of the Weyl fermion is introduced as
follows~[\NIE]:\foot{For definiteness, we will consider the
left-handed Weyl fermion.} The GW chiral matrix is defined
by~$\hat\gamma_5=\gamma_5(1-aD)$.\foot{Note that $(\hat\gamma_5)^2=1$
and~$(\hat\gamma_5)^\dagger=\hat\gamma_5$.} The chiral projectors are
then defined by~$\hat P_\pm=(1\pm\hat\gamma_5)/2$
and~$P_\pm=(1\pm\gamma_5)/2$. Since $P_+D=D\hat P_-$, we can
consistently impose the chirality as $\hat P_-\psi=\psi$
and~$\overline\psi P_+=\overline\psi$. Note that the mass terms
in~\twoxtwo\ are also consistent with this definition of the chirality
because $B\hat P_-=\hat P_-^*B=\hat P_-^TB$.

To define the fermion integration measure
${\rm D}[\psi]{\rm D}[\overline\psi]$ in~\twoxone, one first
introduces basis vectors~$v_j$ ($j=1$, $2$, \dots, $\Tr\hat P_-$),
which satisfy the constraint~$\hat P_-v_j=v_j$
and~$(v_j,v_k)=\delta_{jk}$.\foot{The inner product for spinors is
defined by~$(\psi,\varphi)=a^4\sum_x\psi^\dagger(x)\varphi(x)$.} The
fermion field is then expanded as~$\psi(x)=\sum_jv_j(x)c_j$ and the
measure is defined
by~${\rm D}[\psi]=\prod_j{\rm d}c_j\equiv{\rm d}c_1{\rm d}c_2\cdots
{\rm d}c_{\Tr\hat P_-}$. These conditions, however, do not specify the
measure uniquely; there remains a phase ambiguity that may depend on 
the gauge field. For a different choice of basis
vectors~$\widetilde v_j$, one has
$$
   \widetilde v_j(x)=\sum_kv_k(x)\bigl({\cal Q}^{-1}\bigr)_{kj},
\eqn\twoxfour
$$
with a unitary matrix~${\cal Q}$. The coefficients are thus related
as~$\widetilde c_j=\sum_k{\cal Q}_{jk}c_k$ and the measures differ by
a phase factor,
$\prod_j{\rm d}c_j=\det{\cal Q}\prod_j{\rm d}\widetilde c_j$. How to
choose (and whether it is possible to choose) the phase over the
gauge-field configuration space that is consistent with the gauge
invariance is the central issue in the formulation. The measure for
the anti-fermion is defined similarly but with respect
to~$P_+$ as~${\rm D}[\overline\psi]=\prod_k{\rm d}\overline c_k\equiv
{\rm d}\overline c_1{\rm d}\overline c_2\cdots
{\rm d}\overline c_{\Tr P_+}$, where
$\overline\psi(x)=\sum_k\overline c_k\overline v_k(x)$,
$\overline v_kP_+=\overline v_k$ ($k=1$, $2$, \dots, $\Tr P_+$) and
$(\overline v_k^\dagger,\overline v_l^\dagger)=\delta_{kl}$. The phase
of~${\rm D}[\overline\psi]$ can be chosen as being independent of the
gauge field and it thus has no physical relevance.

An important point to note is that the above construction refers to a
specific topological sector. The number of integration variables
in~${\rm D}[\psi]$ is~$\Tr\hat P_-$, and this number depends on the
gauge-field configuration. In this way, the fermion-number violation
in topologically non-trivial sectors is naturally incorporated. Since
$\Tr\hat P_-$~is an integer~[\HASE], the smoothness of the Dirac
operator in the admissible space~\twoxthree\ guarantees that
$\Tr\hat P_-$~is constant within a connected component in the
admissible space. The full expectation value, including the gauge
field sector, is thus given by
$$
   \VEV{{\cal O}}={1\over{\cal Z}}
   \sum_M\int_M{\rm D}[U]\,e^{-S_{\rm G}}
   {\cal N}(M)e^{i\vartheta(M)}
   \VEV{{\cal O}}_{\rm F}^M,
\eqn\twoxfive
$$
where ${\cal Z}$~is chosen as $\VEV{1}=1$ and $M$~stands for each
connected component in the admissible space. The restriction of the
gauge-field integration to the admissible space may be implemented by
the modified plaquette action~[\LUS] for example. On the other hand,
as already emphasized in~[\LUS], at the moment there is no obvious way
to fix the relative normalization~${\cal N}(M)$ and the relative
phase~$\vartheta(M)$ for different topological sectors. We will come
back to this point in a later section.

\chapter{Global existence of the fermion integration measure}
In this section, for real representations, we will show that it is
possible to construct the gauge-invariant fermion measure globally and
smoothly over the gauge-field configuration space (or more precisely,
within each connected component in the admissible space). The
underlying symplectic structure plays the key role in this.

Take a certain gauge-field configuration~$U(x,\mu)$. We will construct
the basis vectors~$v_j$ introduced in the previous section starting
with a complete set of arbitrarily chosen vectors~$u_j$ in the
constrained space~$\hat P_-u_j=u_j$. We first
set~$v_1=u_1/\sqrt{(u_1,u_1)}$. Next we can take $v_2$
as~$v_2=v_1'\equiv B^{-1}v_1^*$, because
$v_1'$~satisfies $\hat P_-v_1'=v_1'$ and~$(v_1,v_1')=0$,
since~$B^T=-B$ ($v_2$ is correctly normalized, $(v_2,v_2)=1$). Note
that $v_2'=B^{-1}v_2^*=-v_1$. Since $u_j$ span a complete set, we have
$v_2=\sum_{j\neq1}k_ju_j$, where we may assume $k_2\neq0$ without
loss. Thus, we can replace $u_1$ and~$u_2$ in the complete set by
$v_1$ and~$v_2$. Next, we define $v_3$ from~$u_3$ such that it is
orthogonal to $v_1$ and~$v_2$. This can be done by the Gram--Schmidt
method as $\widetilde v_3=u_3-(v_1,u_3)v_1-(v_2,u_3)v_2$
and~$v_3=\widetilde v_3/\sqrt{(\widetilde v_3,\widetilde v_3)}$;
$v_4$~is defined from~$v_3$ by~$v_4=v_3'=B^{-1}v_3^*$. Now we see that
$v_4$~is linearly independent of $v_1$, $v_2$ and~$v_3$, because
$(v_1,v_4)=-(v_3,v_2)=0$, $(v_2,v_4)=(v_3,v_1)=0$ and~$(v_3,v_4)=0$.
Since $v_4=\sum_{j\neq1,2,3}k_j'u_j$, we may replace $u_3$ and (say)
$u_4$ in the complete set by $v_3$ and~$v_4$. Clearly this procedure
can be repeated pairwise and we are left with the orthonormal complete
set~$v_j$ with~$\hat P_-v_j=v_j$ such that $v_{2l}=v_{2l-1}'$
and~$v_{2l-1}=-v_{2l}'$. This basis~$v_j$ can be characterized by
$$
   v_j'(x)=B^{-1}v_j^*(x)=J_{jk}v_k(x),\qquad
   J_{jk}=\delta_{j+1,k}-\delta_{j,k+1},
\eqn\threexone
$$
where $J^\dagger J=1$ and~$J^T=-J$.

For a fixed gauge-field configuration~$U(x,\mu)$, we have shown that
it is always possible to construct $v_j$ such that
$\hat P_-v_j=v_j$, $(v_j,v_k)=\delta_{jk}$ and \threexone\ hold. These
$v_j$ moreover can be {\it smoothly\/} continued to other gauge-field
configurations, at least within a sufficiently small local patch
containing~$U(x,\mu)$. The reason is that the above construction is
purely algebraic and when the gauge field is continuously varied,
$v_j$~changes smoothly. The smoothness of the construction breaks down
only when, for example, $v_2$ happens to have no component of~$u_2$
and we need to change the labelling of~$u_j$'s. But such a situation
cannot occur for sufficiently close neighbors of~$U(x,\mu)$.

Therefore, it is always possible to construct a smooth basis~$v_j$
within a local patch in the gauge-field configuration space such
that~\threexone\ holds. Now we can show that, as long as
condition~\threexone\ is satisfied, the corresponding
measure~${\rm D}[\psi]$---we call this the symplectic measure---is
{\it unique}. The proof of this important fact is simple: assume a
different basis~$\widetilde v_j$ also satisfies~\threexone. Since
$v_j$ and~$\widetilde v_j$ are related by~\twoxfour,
\threexone\ implies that the unitary matrix~${\cal Q}$ satisfies
$J{\cal Q}J^{-1}={\cal Q}^*$, i.e.\ ${\cal Q}$~is symplectic. Namely,
we have $\det{\cal Q}=1$\foot{We define $\xi'\equiv J^{-1}\xi^*$. If
$\xi$~is an eigenvector of~${\cal Q}$, ${\cal Q}\xi=e^{i\theta}\xi$,
then $\xi'$ has the eigenvalue $e^{-i\theta}$. Since $\xi$ and~$\xi'$
are linearly independent and $\xi''=-\xi$, this implies that the
eigenvalues of~${\cal Q}$ always come in pairs as $e^{i\theta}$
and~$e^{-i\theta}$.} and the associated measure for~$v_j$ and that
for~$\widetilde v_j$ are identical.

Now, cover the gauge-field configuration space by a collection of
local coordinate patches. Within each patch, we can construct the
smooth symplectic measure as described above. In an overlap of two
patches, the basis vectors in one patch and that in another patch are
not necessarily the same. However, corresponding {\it measures\/} are
identical, since both are symplectic, and the symplectic measure is
unique. This shows that it is always possible to define a smooth
measure over the gauge-field configuration space. The important point 
is that the construction of the symplectic measure within a local
patch requires only the {\it local\/} information, but nevertheless
the symplectic condition~\threexone\ guarantees the {\it global\/}
consistency of the measure.

Under the infinitesimal variation of the gauge field
$$
   \delta_\eta U(x,\mu)=a\eta_\mu(x)U(x,\mu),
\eqn\threextwo
$$
the measure changes
as~$\delta_\eta{\rm D}[\psi]=-i{\cal L}_\eta{\rm D}[\psi]$, where the
measure term~${\cal L}_\eta$~[\LUS,\LUSC] is defined
by~${\cal L}_\eta=i\sum_j(v_j,\delta_\eta v_j)$. For the symplectic
measure, the measure term identically vanishes,
${\cal L}_\eta=i\sum_l[(v_{2l-1},\delta_\eta v_{2l-1})+
(v_{2l},\delta_\eta v_{2l})]=i\sum_l\delta_\eta(v_{2l-1},v_{2l-1})=0$,
because $v_{2l}=v_{2l-1}'=B^{-1}v_{2l-1}^*$. This implies that the
symplectic measure is independent of the gauge field within a
connected component in the admissible space. Incidentally, since the
measure term transforms
as~$\widetilde{\cal L}_\eta={\cal L}_\eta-i\delta_\eta\ln\det{\cal Q}$
under the change of basis vectors~\twoxfour, {\it any\/} measure with
vanishing measure term~${\cal L}_\eta=0$ is identical to the
symplectic measure up to a constant phase.

It remains to be shown that the symplectic measure is gauge-invariant.
The infinitesimal gauge transformation is given
by~$\eta_\mu(x)=-\nabla_\mu\omega(x)$ in~\threextwo.\foot{%
$\nabla_\mu\omega(x)=[U(x,\mu)\omega(x+a\hat\mu)U(x,\mu)^{-1}-
\omega(x)]/a$ is the covariant difference operator.} By using the
gauge covariance of the Dirac operator~$\delta_\eta D=[R(\omega),D]$
in~\twoxone, we have as the gauge variation
of~$\VEV{{\cal O}}_{\rm F}$,
$$
   \delta_\eta\VEV{{\cal O}}_{\rm F}
   =\VEV{\delta_\eta{\cal O}}_{\rm F}
   +\bigl[\Tr R(\omega)(P_+-\hat P_-)-i{\cal L}_\eta\bigr]
   \VEV{{\cal O}}_{\rm F}.
\eqn\threexthree
$$
In the quantity in square brackets, the first term comes from the
Jacobian of the change of fermion variables and the second term from
the fact that basis vectors themselves change under~\threextwo. We
showed that ${\cal L}_\eta=0$ for the symplectic measure. On the other
hand, noting $P_+^T=BP_+B^{-1}$, $R(\omega)^T=-R(\omega)$ and
$\hat P_-^T=B\hat P_-B^{-1}$ for real representations, we see that the
first term identically vanishes. Namely, expectation values of gauge
invariant operators are always gauge-invariant and the symplectic
measure (or more generally any measure with~${\cal L}_\eta=0$) is
gauge-invariant.\foot{For anomaly-free {\it complex\/}
representations, the quantity~$\Tr R(\omega)(P_+-\hat P_-)$ does not
vanish and the way to (and whether it is possible to) choose
${\cal L}_\eta$ to eliminate the combination inside the square
brackets is the aforementioned problem of the local gauge anomaly.
This problem can be studied by cohomological
techniques~[\LUSCHE,\LUSC,\SUZU]. The current status of our knowledge
concerning~${\cal L}_\eta$ is as follows: when the gauge group
is~$U(1)$, such ${\cal L}_\eta$ has been known non-perturbatively on
finite lattices~[\LUS]. For general compact gauge groups,
${\cal L}_\eta$ has been known, but only to all orders in the
perturbation theory on the infinite lattice~[\SUZU,\LUSCH]. For the
representation in the electroweak $SU(2)\times U(1)$,
${\cal L}_\eta$~has been known non-perturbatively at least on the
infinite lattice~[\KIK].} This establishes the existence of a globally
consistent gauge-invariant measure in any topological sector; there is
no global obstruction for real representations.

\chapter{Fermion expectation values}
In this section, we explicitly compute the expectation
value~\twoxone\ by using the symplectic measure. As shown in the
previous section, the symplectic measure can be constructed starting
with any complete set~$u_j$ satisfying~$\hat P_-u_j=u_j$. A
particularly convenient complete set~$u_j$ is provided by
eigenvectors of the hermitian operator~$D^\dagger D=(\gamma_5D)^2$:
$$
   D^\dagger Du_j(x)=\lambda_j^2u_j(x),\qquad\hat P_-u_j(x)=u_j(x).
\eqn\fourxone
$$
(This choice is analogous to that in the treatment of covariant gauge
anomalies in the continuum theory~[\FUJI].) These two conditions are
consistent because $D^\dagger D$ and~$\hat P_-$ commute. For later
comparison with the Majorana formulation, we need to know some details
concerning the eigenvalue problem~\fourxone. For this, we consider
the auxiliary problem\foot{Note that $\gamma_5D$~is hermitian.}
$$
   \gamma_5D\varphi_n(x)=\lambda_n\varphi_n(x),\qquad
   \hbox{$\lambda_n$: real},\qquad
   n=1,2,\cdots,\Tr1.
\eqn\fourxtwo
$$
The eigenvectors~$\varphi_n$ are classified into three
categories:\foot{Since it is simple to prove the following statements,
we do not give the detailed proof.} (i)~$\lambda_n\neq0$
and~$\lambda_n\neq\pm2/a$. Then $\widetilde\varphi_n\equiv
\gamma_5(1-aD/2)\varphi_n/\sqrt{1-a^2\lambda_n^2/4}$ has the
eigenvalue~$-\lambda_n$; the eigenvalues thus come in pairs as
$\lambda_n$ and~$-\lambda_n$. (ii)~$\lambda_n=\pm2/a$. Denoting
$\Psi_\pm$ as the corresponding eigenvectors, one has
$P_\pm\Psi_\pm=\hat P_\mp\Psi_\pm=\Psi_\pm$. We denote the number
of~$\Psi_\pm$ as~$N_\pm$. (iii)~$\lambda_n=0$. One can choose the
eigenvectors with definite chiralities as
$P_\pm\varphi_0^\pm=\hat P_\pm\varphi_0^\pm=\varphi_0^\pm$. We denote
the number of~$\varphi_0^\pm$ as~$n_\pm$. The number~$n_+-n_-$ is the
analytic index on the lattice~[\HASE], which is constant in a
connected component in the admissible space. For the number of
eigenvectors of the latter two categories, one can show the index
relation~[\CHI]
$$
   n_+-n_-+N_+-N_-=0,
\eqn\fourxthree
$$
starting with~$\Tr\gamma_5=0$. For {\it real\/} representations, all
the eigenvalues {\it including\/} $\lambda_n=0$ and~$\lambda_n=\pm2/a$
are moreover {\it doubly-degenerate}: $\varphi_n$
and~$\varphi_n'=B^{-1}\varphi_n^*$ give the same
eigenvalue~$\lambda_n$, and $\varphi_n'$ is linearly independent
with~$\varphi_n$ because $(\varphi_n,\varphi_n')=0$. In particular,
$N_\pm$ and~$n_\pm$ are even numbers.

Once having obtained the solution of~\fourxtwo, we can obtain all the
solutions of~\fourxone\ by simply multiplying $\hat P_-$, because
$\varphi_n$ span a complete set. In this way, we have: (I)~$u_j$
with $\lambda_j^2\neq0$ and~$\lambda_j^2\neq4/a^2$ from category~(i).
But since $\hat P_-[\widetilde\varphi_n+(1-a\lambda_n/2)\varphi_n]=0$,
only one linear combination of $\varphi_n$ and~$\widetilde\varphi_n$
gives rise to the solution of~\fourxone. Thus the total number of this
type of~$u_j$ is~$[\Tr1-(N_++N_-+n_++n_-)]/2$. (II)~$u_j$
with~$\lambda_j^2=4/a^2$. This is given by~$\hat P_-\Psi_+$ and the
total number is~$N_+$. (III)~$u_j$ with $\lambda_j^2=0$. This is given
by~$\hat P_-\varphi_0^-$ and the total number is~$n_-$.

Following the previous construction from $u_j$ to~$v_j$, we thus
obtain $v_j$ that satisfy \fourxone,
\threexone\ and~$(v_j,v_k)=\delta_{jk}$. Below we will use this
particular basis to compute the expectation value~\twoxone. Recall,
however, that the fermion measure itself is independent of which kind
of basis vectors are employed, as long as the symplectic
condition~\threexone\ is satisfied.

The expectation value~\twoxone\ also depends on how we choose the
phase of~${\rm D}[\overline\psi]$. We fix this phase by the following
natural mapping from $v_j$ to~$\overline v_j$ for~$\lambda_j\neq0$:
$$
   \overline v_j(x)={1\over\lambda_j}v_j^\dagger D^\dagger(x),\qquad
   \lambda_j>0.
\eqn\fourxfour
$$
Note that $\overline v_j$ so constructed satisfies
$\overline v_jP_+=\overline v_j$,
$\overline v_jDD^\dagger=\overline v_j(D\gamma_5)^2=
\lambda_j^2\overline v_j$,
$v_j=D^\dagger\overline v^\dagger_j/\lambda_j$ and
$(\overline v_j^\dagger,\overline v_k^\dagger)=\delta_{jk}$. This
mapping gives rise to the symplectic structure
$\overline v_j'\equiv\overline v_j^*B=J_{jk}\overline v_k$ also for
$\overline v_j$. For the zero modes~$\lambda_j^2=0$, a mismatch
between $\overline v_j$ and $v_j$ may occur and we can take
$\varphi_0^{+\dagger}P_+=\varphi_0^{+\dagger}$ as the basis vectors
for the zero modes in~$\overline v_j$. The total number of these
is~$n_+$.

We have completely fixed the phase ambiguity for the measure
in~\twoxone. What remains to be done is simply the Grassmann
integrals with respect to $c_j$ and~$\overline c_k$. For the partition
function~$\VEV{1}_{\rm F}$, after a careful calculation using the
above relations, we have\foot{In the massless limit~$m\to0$, this
expression may be interpreted
as~$\VEV{1}_{\rm F}=\pm\sqrt{\det\gamma_5D}$, as naively expected for
the Weyl fermion in a real representation. Since eigenvalues
of~$\gamma_5D$ are doubly-degenerate, even if some of the eigenvalues
cross zero according to a deformation of the gauge field, there is no
ambiguity in the sign of the square root~[\WIT] because it is always
an even number of eigenvalues that cross zero~[\HSU]. This explains
(for the vacuum sector) why the Witten anomaly does not appear for
real representations from the viewpoint of the spectral flow.}
$$
\eqalign{
   \VEV{1}_{\rm F}&=
   \prod_{\lambda_n>0\atop\lambda_n\neq2/a}
   \bigl[-(\lambda_n^2+m^2)\bigr]
   \biggl[-\biggl({4\over a^2}+m^2\biggr)\biggr]^{N_+/2}
   (im)^{(n_++n_-)/2}
\cr
   &=i^{[\Tr1-(n_+-n_-)]/2}
   \prod_{\lambda_n>0\atop\lambda_n\neq2/a}
   (\lambda_n^2+m^2)\biggl({4\over a^2}+m^2\biggr)^{N_+/2}
   m^{(n_++n_-)/2},
\cr
}
\eqn\fourxfive
$$
in terms of the eigenvalues~$\lambda_n$ in~\fourxtwo. In this
expression, the product~$\prod_{\lambda_n>0\atop\lambda_n\neq2/a}$ is
understood to be taken {\it without\/} counting the double degeneracy
of $\lambda_n$ (i.e.\ one factor for each~$\lambda_n$). We have used
the index relation~\fourxthree\ in deriving the second line.

The expression~\fourxfive\ holds for any topological sector.
Interestingly, in the massive theory, the partition
function~$\VEV{1}_{\rm F}$ has a definite sign, up to a
proportionality constant that depends only on which topological
sector is concerned through the combination~$n_+-n_-$.\foot{This fact
may be of interest from the viewpoint of numerical simulations.} In
the massless theory~$m\to0$, $\VEV{1}_{\rm F}$ vanishes when there
exists a zero mode, as should be the case. The general fermion
expectation value~$\VEV{{\cal O}}_{\rm F}$ is computed as usual by
$\VEV{1}_{\rm F}$ times the Wick contractions of fermion fields. The
basic contractions are given by
$$
\eqalign{
   &{\VEV{\psi(x)\overline\psi(y)}_{\rm F}\over
   \VEV{1}_{\rm F}}
   =\hat P_-{1\over D^\dagger D+m^2}D^\dagger P_+(x,y),
\cr
   &{\VEV{\psi(x)\psi^T(y)}_{\rm F}\over
   \VEV{1}_{\rm F}}
   =\hat P_-{-im\over D^\dagger D+m^2}B^{-1}\hat P_-^T(x,y),
\cr
   &{\VEV{\overline\psi^T(x)\overline\psi(y)}_{\rm F}\over
   \VEV{1}_{\rm F}}
   =P_+^TB{im\over DD^\dagger+m^2}P_+(x,y).
\cr
}
\eqn\fourxsix
$$
It is easy to express these basic contractions in terms of the
eigenvalues and eigenfunctions in~\fourxtwo\ by noting
$\hat P_-(x,y)=\sum_jv_j(x)v_j^\dagger(y)$~etc., although we do not
write them down explicitly. For example, in the massless
limit~$m\to0$, $\VEV{\psi(x)\psi^T(y)}_F/\VEV{1}_F\to
-i\varphi_0^-(x)\varphi_0^{-\dagger}(y)B^{-1}/m$ and it thus precisely
cancels one $m$ in~\fourxfive\ due to one pair of left-handed zero
modes. In this way, any fermion expectation value in any topological
sector is obtained by combining \fourxfive\ and~\fourxsix. Note that,
according to the above expressions, expectation values of
gauge-invariant operators are manifestly gauge-invariant.

\chapter{Matching to the Majorana formulation}
As noted in the introduction, in four-dimensional continuum
(unregularized) theories, the Weyl fermion in the real representation
is equivalent to the Majorana fermion. Thus it is of interest to see
how this equivalence is realized in the present formulation in which
the left-right chiralities are treated asymmetrically. The lattice
implementation of the Majorana (left--right symmetric) fermion would
be given by
$$
   S_{\rm F}^{\rm Majorana}=a^4\sum_x\biggl[
   {1\over2}\chi^T(x)CD\chi(x)+{1\over2}im\chi^T(x)C\gamma_5\chi(x)
   \biggr],
\eqn\fivexone
$$
where $\chi$~is a four-component {\it unconstrained\/} spinor field.
Note that $(CD)^T=-CD$ and~$(C\gamma_5)^T=-C\gamma_5$ being consistent
with the Fermi statistics and that the mass term is gauge-invariant
for real representations. The expectation value is then given by
$\VEV{{\cal O}}_{\rm F}^{\rm Majorana}=
\int{\rm D}[\chi]\,{\cal O}e^{-S_{\rm F}^{\rm Majorana}}$, where the
fermion integration measure is defined
by~$\chi(x)=\sum_n\varphi_n(x)b_n$ ($\varphi_n$'s are certain
orthonormal basis vectors)
and~${\rm D}[\chi]=\prod_n{\rm d}b_n\equiv{\rm d}b_1{\rm d}b_2\cdots
{\rm d}b_{\Tr 1}$. The important difference from the Weyl formulation
is that the Majorana formulation can be set up without referring to a
particular topological sector, because the number of integration
variables is always the same. Namely, the above definition is
uniform for all topological sectors.\foot{This is analogous to the
situation for the Dirac fermion in lattice QCD in which one usually
never worries about the relative weight for the fermion measure in
different topological sectors.} This property of the Majorana
formulation has an interesting implication, as we will discuss in the
next section.

We can take the eigenvectors in~\fourxtwo\ as the basis
vectors~$\varphi_n$. With this choice, we obtain as the fermion
partition function
$$
\eqalign{
   &\VEV{1}_{\rm F}^{\rm Majorana}
   =\prod_{\lambda_n>0\atop\lambda_n\neq2/a}(\lambda_n^2+m^2)
   \biggl(-{2\over a}-im\biggr)^{N_+/2}
   \biggl({2\over a}-im\biggr)^{N_-/2}
   (-im)^{(n_++n_-)/2}
\cr
   &=-i^{[-(n_+-n_-)]/2}\biggl({2\over a}-im\biggr)^{(n_+-n_-)/2}
   \prod_{\lambda_n>0\atop\lambda_n\neq2/a}
   (\lambda_n^2+m^2)
   \biggl({4\over a^2}+m^2\biggr)^{N_+/2}
   m^{(n_++n_-)/2},
\cr
}
\eqn\fivextwo
$$
where from the first line to the second line we have used
\fourxthree\ and the fact that $n_-$~is an even number. Note that
$N_\pm$ appear symmetrically in the first expression, because of the
left--right-symmetric treatment in the Majorana formulation.

From~\fivexone, the fermion partition function in the Majorana
formulation is given by the
Pfaffian~$\VEV{1}_{\rm F}^{\rm Majorana}\propto\Pf(CD+imC\gamma_5)$
and \fivextwo~gives the precise meaning of this Pfaffian. In the
massless limit, $\VEV{1}_{\rm F}^{\rm Majorana}\propto\Pf CD$
and, when the overlap-Dirac operator~[\NEUBER] is employed as~$D$,
this coincides with the expression in~[\NEUBE], which is based on a
factorization property of the domain wall~[\KAPL,\KAP] (with the
infinite five-dimensional separation) or the overlap~[\NAR] fermion
determinant in vector-like theories. In this limit, \fivextwo\ reduces
to~$\VEV{1}_{\rm F}^{\rm Majorana}=
-\prod_{\lambda_n>0\atop\lambda_n\neq2/a}\lambda_n^2(4/a^2)^{N_+/2}$
(assuming there is no zero mode), which manifestly has a definite
sign. This is important from the viewpoint of numerical
simulations~[\HSU,\KAP,\CUR--\NEUBE], because the fermion partition
function then allows a statistical weight interpretation. This
property with the overlap-Dirac operator has been shown~[\NEUBE] by
appealing to the limiting procedure from the domain wall fermion with
finite five-dimensional separation. Here we have shown the same
property by using general properties of the GW Dirac operator alone.

Comparing \fivextwo\ and~\fourxfive, we find
$$
   \VEV{1}_{\rm F}^{\rm Majorana}=-(-1)^{\Tr1/4}
   \biggl({2\over a}-im\biggr)^{(n_+-n_-)/2}
   \VEV{1}_{\rm F}^{\rm Weyl}.
\eqn\fivexthree
$$
Namely, two formulations match up to a proportionality constant that
depends only on the topological sector. If one is concerned with a
particular topological sector, two formulations are therefore
completely equivalent. For the basic contraction, we find
$$
\eqalign{
   {\VEV{\chi(x)\chi^T(y)}_{\rm F}^{\rm Majorana}\over
    \VEV{1}_{\rm F}^{\rm Majonara}}
   &={1\over(\gamma_5D)^2+m^2}(D^\dagger-im\gamma_5)C^{-1}(x,y)
\cr
   &={2/a\over2/a-im}
   {\VEV{\Bigl[\psi(x)-C^{-1}\overline\psi^T(x)\Bigr]
   \Bigl[\psi^T(y)-\overline\psi(y)C^{-1T}\Bigr]
   }_{\rm F}^{\rm Weyl}\over\VEV{1}_{\rm F}^{\rm Weyl}}
\cr
   &\qquad
   -{1\over2/a-im}\gamma_5C^{-1}a^{-4}\delta_{x,y},
\cr
}
\eqn\fivexfour
$$
where, in deriving the last expression, we have
noted~$1=P_++\hat P_--a\gamma_5D/2$. The relation in the opposite
direction is given by
$$
\eqalign{
   &{\VEV{\psi(x)\overline\psi(y)}_{\rm F}^{\rm Weyl}
   \over\VEV{1}_{\rm F}^{\rm Weyl}}
   =-{2/a\over2/a-im}
   \hat P_-
   {\VEV{\chi(x)\chi^T(y)}_{\rm F}^{\rm Majorana}\over
    \VEV{1}_{\rm F}^{\rm Majonara}}C^TP_+
\cr
   &{\VEV{\psi(x)\psi^T(y)}_{\rm F}^{\rm Weyl}\over
   \VEV{1}_{\rm F}^{\rm Weyl}}
   ={2/a\over2/a-im}
   \left[\hat P_-
   {\VEV{\chi(x)\chi^T(y)}_{\rm F}^{\rm Majorana}\over
    \VEV{1}_{\rm F}^{\rm Majonara}}\hat P_-^T
   -{a\over2}\hat P_-(x,y)\gamma_5C^{-1}\right]
\cr
   &{\VEV{\overline\psi^T(x)\overline\psi(y)}_{\rm F}^{\rm Weyl}\over
   \VEV{1}_{\rm F}^{\rm Weyl}}
   ={2/a\over2/a-im}
   \left[P_+^TC
   {\VEV{\chi(x)\chi^T(y)}_{\rm F}^{\rm Majorana}\over
    \VEV{1}_{\rm F}^{\rm Majonara}}C^TP_+
   +{a\over2}CP_+a^{-4}\delta_{x,y}\right].
\cr
}
\eqn\fivexfive
$$
Therefore, with these rules \fivexfour\ and~\fivexfive, the
expectation values are identical in the two formulations, up to
contact terms.\foot{Since the kernel~$\hat P_-(x,y)$ decays
exponentially with a fixed range in the lattice units~[\HER], this can
effectively be regarded as a contact term in the continuum limit.} In
particular, they lead to the same physical amplitudes with that
matching rule.

\chapter{Relative normalization for different topological sectors}
We have seen that there is a complete matching between the Weyl
formulation and the Majorana formulation for real representations. In
this section, we present a possible implication of this matching for
the relative normalization of the fermion measure in different
topological sectors (the factor~${\cal N}(M)$ in~\twoxfive) for the
Weyl fermion belonging to the {\it complex\/} representation. For
complex representations, the mass term breaks the gauge symmetry. We
thus restrict our problem to the massless theory.

Suppose that we have a consistent gauge-invariant measure for the Weyl
fermion belonging to the {\it complex\/} representation, which is
specified by the basis vectors~$v_j$. Then the set of
vectors~$B^{-1}v_j^*$ naturally provides a consistent gauge-invariant
measure for the complex conjugate representation~$R^*$. With this
choice of measure for~$R^*$, we have
$\VEV{{\cal O}}_{{\rm F},R}^*=\VEV{{\cal O}^*}_{{\rm F},R^*}$ and, as
naively expected~[\ALV],
$$
\eqalign{
   \bigl|\VEV{{\cal O}}_{{\rm F},R}\bigr|^2
   &=\VEV{{\cal O}^*}_{{\rm F},R^*}\VEV{{\cal O}}_{{\rm F},R}
\cr
   &=\VEV{{\cal O}^*{\cal O}}_{{\rm F},R\oplus R^*},
\cr
}
\eqn\sixxone
$$
where the measure for the {\it real\/} representation~$R\oplus R^*$ is
specified by the basis vectors
$V_{2l-1}=(v_l,0)^T$ and~$V_{2l}=(0,B^{-1}v_l^*)^T$. This measure is
symplectic with respect
to~$V_j'\equiv\Bigl({0\,\,1\atop1\,\,0}\Bigr)B^{-1}V_j^*$ and we can
thus apply the previous arguments. In particular, \sixxone\ shows that
we can compute the modulus of~$\VEV{{\cal O}}_{{\rm F},R}$ by using
\fourxfive\ and~\fourxsix\ with~$m\to0$.

For the real representation~$R\oplus R^*$, we may use also the
Majorana formulation. From \sixxone, \fivexthree\ and~\fivexfive, we
know that for a fixed topological sector:\foot{Although the
{\it phase\/} of the proportionality constant in this expression
depends on a way we specified the phase of~${\rm D}[\chi]$, this does
not affect the following argument for the
{\it normalization}~${\cal N}(M)$.}

$$
   \bigl|\VEV{{\cal O}}_{{\rm F},R}\bigr|^2
   =-(-1)^{\Tr1/4}
   \biggl({a\over2}\biggr)^{n_+-n_-}
   \VEV{{\cal O}^*{\cal O}}_{{\rm F},R\oplus R^*}^{\rm Majorana},
\eqn\sixxtwo
$$
up to contact terms.\foot{Eq.~\fivexfive\ shows that the substitution
rule from the Weyl formulation to the Majorana formulation is given by
$\psi(x)\to\hat P_-\chi(x)$ and~$\overline\psi(x)\to-\chi^T(x)C^TP_+$.
} In this expression, $n_\pm$~refer to the numbers of zero modes of
the {\it original\/} Weyl fermion in the complex representation~$R$.
Now, as already emphasized, the Majorana formulation is
{\it uniform\/} for all topological sectors. Thus it is quite natural
to adjust the normalization of~$\VEV{{\cal O}}_{\rm F,R}$ as it
coincides with the normalization of the Majorana formulation for all
topological sectors. Namely, we may define the relative weight for a
topological sector as
$$
   \VEV{{\cal O}}_{\rm F}
   \rightarrow
   \biggl({2\over a}\biggr)^{(n_+-n_-)/2}e^{i\vartheta}
   \VEV{{\cal O}}_{\rm F}.
\eqn\sixxthree
$$
The natural prescription for the full expectation value would thus be
$$
   \VEV{{\cal O}}={1\over{\cal Z}}
   \sum_M\int_M{\rm D}[U]\,e^{-S_{\rm G}}
   \biggl({2\over a}\biggr)^{(n_+-n_-)/2}e^{i\vartheta(M)}
   \VEV{{\cal O}}_{\rm F}^M,
\eqn\sixxfour
$$
where the relative phase~$\vartheta(M)$ cannot be fixed from the
present argument. Assuming that the operator~${\cal O}$ has a
definite mass dimension, the dimensionful
factor~$(1/a)^{(n_+-n_-)/2}$ compensates changes of the mass dimension
of~$\VEV{{\cal O}}_{\rm F}^M$, which depends
on~$\Tr(P_+-\hat P_-)=n_+-n_-$ (note that the mass dimension of the
Grassmann integration~${\rm d}c_j$ is~$1/2$). This is a natural
requirement for~${\cal N}(M)$. On the other hand, the relative
normalization~$2^{(n_+-n_-)/2}$ was determined from the matching with
the Majorana formulation. If one chooses the normalization of the GW
relation as~$D\gamma_5+\gamma_5D=kaD\gamma_5D$, the number~$2$ changes
to~$2/k$. Therefore, the Weyl formulation will automatically give rise
to the natural relative normalization by choosing the normalization
of the Dirac operator as~$k=2$.

\chapter{Conclusion}
The real representation, owing to its simplicity with regard to the
local gauge anomaly, provides an interesting example with which one
can work out all the quantities in L\"uscher's gauge-invariant lattice
formulation. We hope that we clearly illustrated some global issues
in the formulation with this simple example. An interesting
implication of the present analysis is that the matching to the
Majorana formulation provides a natural normalization of the
fermion-integration measure in different topological sectors. This
could be physically relevant, for example, when considering the
absolute magnitude of fermion-number-violating processes in chiral
gauge theories.

The question raised by Taku Izubuchi many years ago initially
motivated the present work. I am indebted to Yoshio Kikukawa for
enjoyable discussions and to Martin L\"uscher for helpful discussions
and suggestions, which quite enriched the contents of this paper. I~am
grateful to members of the CERN Theory Division, where this work was
done, especially Patricia Ball, Pilar Hern\'andez, Karl Jansen and
Hartmut Wittig for their kind help.
\refout
\bye